\documentclass[reprint,
 superscriptaddress,
 amsmath,amssymb,
 aps,
 prx,
 showkeys
]{revtex4-2}
\usepackage[sort&compress]{natbib}
\usepackage[utf8]{inputenc}
\usepackage{graphicx}
\usepackage{dcolumn}
\usepackage{braket}
\usepackage{setspace}
\usepackage{bm}
\usepackage{amsmath}
\usepackage{csquotes}
\usepackage{mathrsfs}
\usepackage{titlesec}
\usepackage{subfigure}
\usepackage[svgnames]{xcolor}
\usepackage[colorlinks,citecolor=blue, urlcolor=blue,bookmarks=false,hypertexnames=true]{hyperref}
\usepackage{soul}
\providecommand{\abs}[1]{\lvert#1\rvert}

\begin{document}

\title{Smart Machine Vision for Universal Spatial Mode Reconstruction}

\author{José D. Huerta-Morales}
\affiliation{Instituto de Ciencias Nucleares, Universidad Nacional Aut\'onoma de M\'exico, Apartado Postal 70-543, 04510 Cd. Mx., M\'exico}

\author{Chenglong You}
\affiliation{Quantum Photonics Laboratory, Department of Physics \& Astronomy, Louisiana State University, Baton Rouge, LA 70803, USA}

\author{Omar S. Maga\~na-Loaiza}
\affiliation{Quantum Photonics Laboratory, Department of Physics \& Astronomy, Louisiana State University, Baton Rouge, LA 70803, USA}

\author{\\ Shi-Hai Dong}
\email{dongsh2@yahoo.com}
\affiliation{Research Center for Quantum Physics, Huzhou University, Huzhou 313000, China}
\affiliation{Laboratorio de Ciencias de la Información Cuántica, CIC, Instituto Politécnico Nacional, UPALM, C.P 07700, CDMX, México}

\author{Roberto de J. Le\'on-Montiel}
\email{roberto.leon@nucleares.unam.mx}
\affiliation{Instituto de Ciencias Nucleares, Universidad Nacional Aut\'onoma de M\'exico, Apartado Postal 70-543, 04510 Cd. Mx., M\'exico}

\author{Mario A. Quiroz-Ju\'{a}rez}
\email{maqj@fata.unam.mx}
\affiliation{Centro de F\'{i}sica Aplicada y Tecnolog\'{i}a Avanzada, Universidad Nacional Aut\'onoma de M\'exico, Boulevard Juriquilla 3001, 76230 Quer\'{e}taro, M\'exico}



\date{\today}

\begin{abstract}
Structured light beams, in particular those carrying orbital angular momentum (OAM), have gained a lot of attention due to their potential for enlarging the transmission capabilities of communication systems. However, the use of OAM-carrying light in communications faces two major problems, namely distortions introduced during propagation in disordered media, such as the atmosphere or optical fibers, and the large divergence that high-order OAM modes experience. While the use of non-orthogonal modes may offer a way to circumvent the divergence of high-order OAM fields, artificial intelligence (AI) algorithms have shown promise for solving the mode-distortion issue. Unfortunately, current AI-based algorithms make use of large-amount data-handling protocols that generally lead to large processing time and high power consumption. Here we show that a low-power, low-cost image sensor can itself act as an artificial neural network that simultaneously detects and reconstructs distorted OAM-carrying beams. We demonstrate the capabilities of our device by reconstructing (with a 95$\%$ efficiency) individual Vortex, Laguerre-Gaussian (LG) and Bessel modes, as well as hybrid (non-orthogonal) coherent superpositions of such modes. Our work provides a potentially useful basis for the development of low-power-consumption, light-based communication devices.

\end{abstract}

\maketitle

\section{Introduction} 

Various families of structured light beams with OAM have drawn significant attention, mainly because they have emerged as a highly promising avenue for enhancing the capacity and robustness of communication systems \cite{wang2012terabit, willner2015optical, mirhosseini2015high, krenn2016twisted, liu2021non,bhusal2021}. Notably, the high dimensionality of OAM beams is currently utilized by optical communication systems to encode multiple bits of information in a single photon. This makes them especially promising for increasing data-transmission capabilities \cite{yao2011orbital, mirhosseini2015high, padgett2017orbital, magana2019quantum, lollie2022high}. 
Theoretically speaking, the number of OAM modes that can be encoded in a single photon is not fundamentally restricted \cite{mirhosseini2016wigner}. Nevertheless, in practical applications, it has been shown that the number of supported OAM modes can be severely restricted to below sixty \cite{wang2015ultra, wan2022divergence}. Interestingly, to ensure an adequate level of optical power on the receiver's side, the number often remains below twenty \cite{lei2015massive, krenn2014communication, wang2011high, ren2016experimental1}. It is worth pointing out that this limitation arises from the rapid angle divergence of the beam, as the OAM order increases, which ultimately leads to significant power loss \cite{wan2022divergence, padgett2015divergence, zhang2021realization}. Moreover, higher-order OAM modes are known to be more susceptible to turbulence as compared to lower-order modes, so they can experience nontrivial channel cross-talk \cite{wang2022deep}. To overcome the beam-divergence problem, one option is to consider different degrees of freedom of structured light. Specifically, one can prepare superpositions of LG modes with different radial and azimuthal components \cite{raskatla2022speckle}. Another option is to work with superpositions of non-orthogonal modes. This approach exhibits a constant variation in the diameter of the beam spot, mitigating the issue of rapid divergence \cite{wan2022divergence, chen2021uca}.

The detection and reconstruction of OAM modes modified by propagation in disordered systems, such as the atmosphere or optical fibers, constitutes one of the major challenges in the implementation of OAM-based communication protocols \cite{mair2001entanglement, zambrini2006quasi, pires2010measurement, malik2012measurement1, zhou2017orbital}. Traditional approaches for the detection and reconstruction of OAM spectra rely on complex interferometric setups, projective measurements, or phase retrieval algorithms \cite{pires2010measurement, berkhout2010efficient, mirhosseini2013efficient, choudhary2018measurement, gong2019optical, ding2022spatial}. These methods often require a number of measurements that scale with the number of encoded modes in the input beam. Furthermore, some of them demand precise alignment and calibration \cite{pires2010measurement, karimi2009efficient, berkhout2010efficient, lavery2012refractive, mirhosseini2013efficient, gibson2004free, kulkarni2017single, zhou2017sorting}. Notably, AI-based algorithms have progressively revolutionized a wide range of disciplines, and optics is no exception. Numerous applications can be found in imaging \cite{lyu2017deep, bhusal2022smart, lim2020three}, ultrafast optics \cite{genty2021machine}, topology \cite{pilozzi2018machine}, optical metrology \cite{you2020identification, feng2019fringe}, and biomedical diagnosis \cite{shen2017deep, suzuki2017overview, quiroz2021identification}. In particular, deep learning approaches have demonstrated promising results for solving detection, reconstruction, and sorting of spatial modes where degenerate intensity distributions exhibit different OAM spectra \cite{raskatla2021deep, raskatla2021deep2, raskatla2022speckle, liu2019superhigh, wang2019convolutional, xiong2020convolutional, giordani2020machine, wang2021deep, wang2022deep}. Unfortunately, most of these strategies make use of highly-dense convolutional layer models, combined with transfer-learning methods, which demand high power consumption and computational resources. For this reason, the detection and reconstruction of the OAM spectra are often carried out off-line in computers with sophisticated hardware. Note that the development of portable-integrated electronic devices faces significant constraints due to the computing power required to execute these algorithms \cite{wang2023intelligent, vieira2022low}.


\begin{figure*}
\includegraphics[width=1\linewidth]{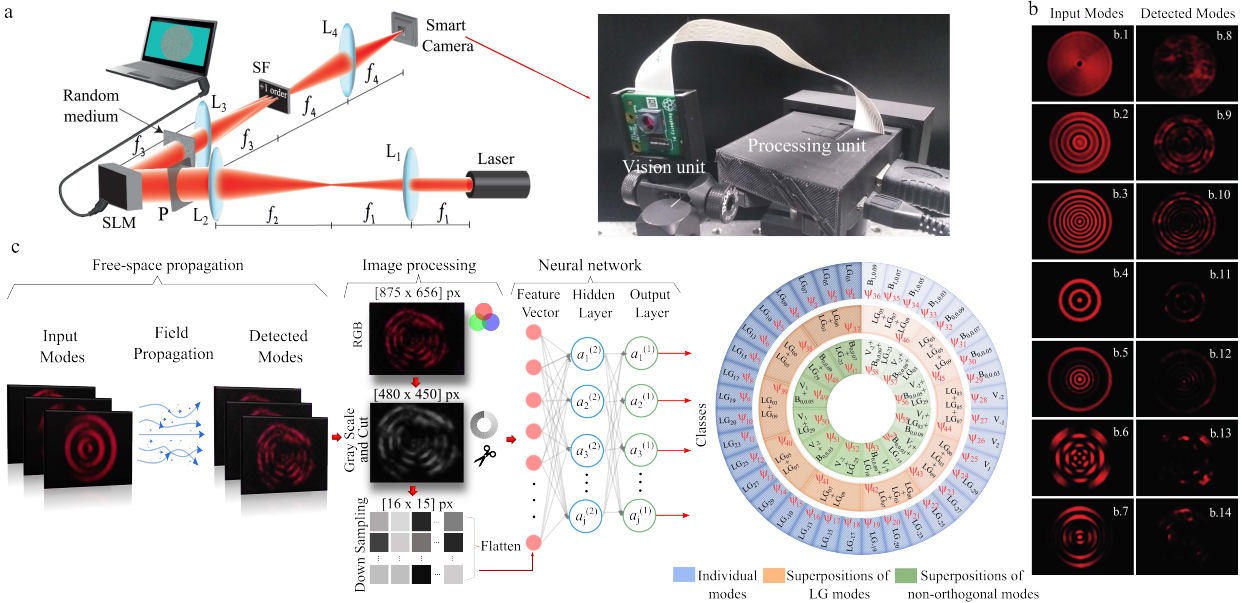}
\caption{\label{fig:Arreglo} \textbf{Generation of individual and hybrid superpositions of OAM-carrying light modes.} \textbf{a} Schematic representation of the experimental setup. $f_{j}$, with $j=1,2,3,4$ stands for the focal lengths of lenses L1, L2, L3, and L4, respectively. P is a linear polarizer, and SF corresponds to the spatial filter used to isolate the +1 diffraction order, produced by the hologram displayed on the spatial light modulator (SLM), see main text for details. The inset on the right hand shows a photograph of our smart machine-vision device. \textbf{b} Experimentally-generated spatial modes of light. The left column (b.1-b.7) corresponds to the intensity distribution of distortion-free OAM-carrying modes. The modes on the right column (b.8-b.14) shows the effect of propagation through ground glass. The individual (single) modes shown in the figure are the following: (b.1,b.8) $V_{-2}$, (b.2,b.9) $B_{0,0.03}$, and (b.3,b.10) $LG_{07}$. The orthogonal coherent superpositions are given in (b.4,b.11) $LG_{00}+LG_{05}$ and (b.5,b.12) $LG_{05}+LG_{07}+LG_{09}$, whereas the non-orthogonal superpositions are shown in (b.6,b.13) $V_{-2}+LG_{-25}$ and (b.7,b.14) $V_{-2}+B_{0,0.09}+LG_{03}$. \textbf{c} Diagrammatic representation of our demultiplexing protocol. Spatial modes propagate in free space, where distortions are induced by means of a ground glass. Distorted modes are then collected by our smart camera and subjected to a data-processing protocol that generates an input vector for an artificial neural network. Three neural networks have been trained for different purposes: (i) to identify individual modes (classes displayed in the outer, blue circle), (ii) to reconstruct superpositions of LG modes whose classes are presented in the middle, orange circle, and (iii) to recover individual modes (outer blue circle), superpositions of LG modes (middle orange circle), and superpositions of non-orthogonal modes (inner green circle). Note that the last neural network includes the classes shown in the blue, orange, and green circles.} 
\end{figure*}

In this work, we demonstrate that a low-power, low-cost image sensor can behave itself as an artificial neural network, capable of detecting and reconstructing distorted OAM-carrying optical beams. This device constitutes the building block of an integrated electronic system that can accurately identify spatial modes, embedded in complex superpositions, by exploiting the self-learning properties of advanced machine learning algorithms. Our methodology involves training of the artificial neural networks on a dataset comprising experimental spatial-mode intensity distributions that propagate through ground glass. The ground glass is introduced to distort and modify the original light-beam profiles. We show the effectiveness of our approach on various types of beams, including Bessel, Vortex, and LG beams. Remarkably, the results demonstrate that our smart machine-vision device can reconstruct OAM modes, and their hybrid (non-orthogonal) superpositions with 95$\%$ efficiency. 

The paper is structured as follows. Section II describes the generating functions of spatially structured light modes. Specifically, we provide the description for Bessel, Vortex, and LG beams. Section III presents the detailed description of the experimental setup used in this work. In Section IV, we demonstrate the capabilities of our smart device for the detection and reconstruction of OAM modes, as well as their superpositions, and Section V is devoted to our conclusions.

\section{Theoretical Framework}
\label{sec:model}
Let us start by describing the generating functions of the spatially-structured light modes used in this work. These are experimentally produced by shaping a laser beam with computer-generated holograms, displayed in a phase-only spatial light modulator (SLM). The first family of modes that we consider belongs to the well-known Vortex beams. These are characterized by the azimuthal phase $V_{l}=\exp(-i l \phi)$, where the topological charge $l$ is related to the OAM value of the beam. Its magnitude determines the order of the helical phase \cite{PhotonRes.4.2016,PhysRevA.98.043846}. The second family of beams used in this work corresponds to the LG beams, whose general mathematical expression, written in cylindrical coordinates at $z=0$, is given by \cite{J.Opt.Soc.Am.A.28.2253,PhysRevA.92.063841,Opt.Express.25.17382}, 
\begin{eqnarray}
    LG_{lp}(r,\phi,0) = &&\sqrt{ \frac{2p!}{\pi \left( p + \abs{l}  \right)! }} \left( \frac{1}{\omega_{0}} \right) \left( \frac{\sqrt{2} r}{\omega_{0}}  \right)^{\abs{l}}  \nonumber \\ & & \times L_{p}^{\abs{l}} \left( \frac{2r^{2}} {\omega_{0}^{2}} \right) \exp \left[il\phi - \frac{r^2}{\omega_{0}^2}\right],
\end{eqnarray}
where $L_{p}^{\abs{l}}$ is the associated Laguerre polynomial, $\omega_{0}$ is beam waist radius, and $l$, $p$ are the OAM and the radial mode numbers, respectively. For the sake of completeness, we provide the definition of $r=\sqrt{x^2+y^2}$ and $\phi=\arctan{\left( y/x \right)}$ in Cartesian coordinates. Note that the associated Laguerre polynomial of order $p$ and degree $\abs{l}$ can be expanded as
\begin{eqnarray}
    L_{p}^{\abs{l}} \left( \frac{2r^{2}} {\omega_{0}^{2}} \right) = \sum_{m=0}^{p} \frac{(-1)^{m}}{m!} \binom{p+\abs{l}}{p-m} \left( \frac{2r^{2}} {\omega_{0}^{2}} \right)^{m}.
\end{eqnarray}
The third family comprises the so-called $l$th-order Bessel beams, which can mathematically be written as \cite{Appl.Opt.52.4566,Opt.Lett.40.3739,J.Opt.Soc.Am.A.28.2018},
\begin{eqnarray}  \label{eq:Bessel_Beam}
B_{l,\kappa_{r}}(r,\phi,0) = J_{l}(\kappa_{r} r) \exp(il\phi),
\end{eqnarray}
where $J_{l}$ stands for the Bessel function of the first kind, whereas $\kappa_{r}$ is the radial wave vector, i.e., the wave vector in the transverse direction, which determines the effective range of the beam or, alternatively, the ring spacing of the Bessel beam \cite{Appl.Opt.54.8030}. Note that the expression for the Bessel beam [Eq. (\ref{eq:Bessel_Beam})] is also described in cylindrical coordinates at $z=0$. The amplitude of the field is controlled by $J_{l}(\kappa_{r} r)$, while $\exp(il\phi)$ represents the azimuthal phase. 

\section{Generation of OAM Modes }
\label{sec:experimental_setup}

The experimental setup for the generation of different OAM-carrying modes is shown in Figure~\ref{fig:Arreglo}a. For the light source, we make use of a fiber-coupled laser diode at a wavelength of $660$~nm. The laser beam is collimated by means of a single-mode fiber, and a fiber collimation package. We magnify the beam spot-size using a basic Keplerian telescope comprising a plano-convex lens (L1) with focal distance $f_{1}=75$~mm and a bi-convex lens (L2) with $f_{2}=500$~mm. The phase modulation of the light is done by using a PLUTO-2-NIR-011 Phase-Only SLM. The SLM modulates light carrying a specific linear polarization, thus we place a linear polarizer (P) between L2 and the SLM.  

The computer-generated holograms displayed on the SLM are prepared using the well-known Kinoform approach \cite{book_goodman}, where the complex fields to be reconstructed are combined with a reference wave of the form $\psi_{R}(x) = \exp\left( - i 2\pi x /p \right)$ with $p$ being the period of the reference wave's linear phase. This configuration allows us to encode the Fourier transform of the complex field in the $+1$ diffraction order of the grating produced by the reference wave. The diffraction order is then isolated by means of a spatial filter (SF) located at the back focal plane of a bi-convex lens (L3) with focal length $f_{3}=400$~mm. We perform a second optical Fourier transform, and thus recover the sought-after complex field, by using a third bi-convex lens (L4) with a focal length of $f_{4}=200$~mm. Note that distortions in the spatial modes are introduced by means of a ground glass placed between the SLM and the L3 lens.


The generated optical modes are captured by our smart camera. The device comprises a low-cost Raspberry Pi camera-module connected to a processing-unit Raspberry Pi 3 Model B via a camera serial interface port [see the photograph on the right-hand side of Fig.~\ref{fig:Arreglo}a]. The Raspberry Pi 3 board is controlled using Python 3 programming language. The technical details of the programming of the device can be found in the Appendix A. To acquire the intensity distributions of the modes, we remove the IR filter from the back-illuminated CMOS. The vision unit adopts the Sony IMX708 Image Sensor, which has the capability to capture high-resolution 12-megapixel images at a resolution of 4608 × 2592 pixels. The sensor has a diagonal size of 7.4 mm and pixel size of 1.4 $\mu$m × 1.4 $\mu$m. It is worth pointing out that, for the sake of simplicity---and to ensure efficient handling of the dataset in a resource-limited device---we reduce the acquisition resolution to 875 x 656 pixels. This reduction allows for easier operation and management of the image dataset without compromising the quality of the captured images.  
Some examples of experimentally-produced spatial modes of light, including superpositions between orthogonal and non-orthogonal modes, are visualized in Fig.~\ref{fig:Arreglo}b. The left column shows the pristine, unaffected OAM modes; whereas the right column shows the result of their propagation through the ground glass.


\section{Results and Discussion}

To identify individual OAM-carrying modes, and their corresponding orthogonal and non-orthogonal coherent superpositions, our smart machine-vision device exploits the self-learning properties of artificial neural networks. Our neural networks consist of sigmoid neurons in a single hidden layer and softmax neurons in the output layer. These are optimized using the scaled conjugate gradient back-propagation algorithm \cite{moller1993scaled}. The optimization process aims to minimize the cross-entropy \cite{shore1981properties, de2005tutorial}, which serves as the loss function, as it has demonstrated suitability for classification tasks \cite{shore1981properties}. From the dataset, we allocate 70\% of the dataset for training, with 15\% reserved for validation and an additional 15\% for testing. To avoid bias in identification, all networks were trained and tested with balanced data, and the testing data was excluded from the training stage.

Our models are trained using the intensity distributions of experimentally detected OAM modes that propagate in free space, as well as in ground glass. We have prepared various types of beams, such as LG beams with different values of ($l$, $p$), Bessel beams with varying orders of $l$, and the radial wave vector $\kappa_{r}$, and Vortex beams with different $l$ values. Examples of such modes are depicted in the first three rows of Fig. \ref{fig:Arreglo}b. For the sake of completeness, we select some example superpositions, including two and three LG beams, and two and three non-orthogonal modes comprising Bessel, Vortex, and LG beams. These are displayed in the last four rows of Fig. \ref{fig:Arreglo}b. Appendix B shows the complete, larger catalog of spatial modes (see Fig. \ref{fig:One_Mode}) that our trained neural networks can recognize. For each spatial mode from the catalog, we collect one hundred observations. Note that we have intentionally introduced distortions to the beams using a movable ground glass. Consequently, the detected mode exhibits a different intensity distribution in each observation, thus allowing the neural network to \emph{learn} all possible distortions that a specific OAM-carrying mode may experience during propagation through the ground glass. The distorted modes collected by our smart camera undergo a processing stage that includes RGB to grayscale conversion, image cropping, and downsampling (see Fig. \ref{fig:Arreglo}c). The conversion of RGB to grayscale enables the elimination of hue and saturation information, while retaining luminance. After the conversion, images are cropped to remove black edges. The original resolution of 875 x 656 pixels is reduced to 480 x 450 pixels after this step. To further decrease the data dimension, a down-sampling process is applied to the resulting monochromatic images by averaging clusters of 30 x 30 pixels, resulting in images of 16 x 15 pixels. Finally, the low-resolution image is rearranged as a column vector and used as the input vector for the multilayer neural network as shown in Fig. \ref{fig:Arreglo}c.

\begin{figure}[t!]
\includegraphics[width=1\linewidth]{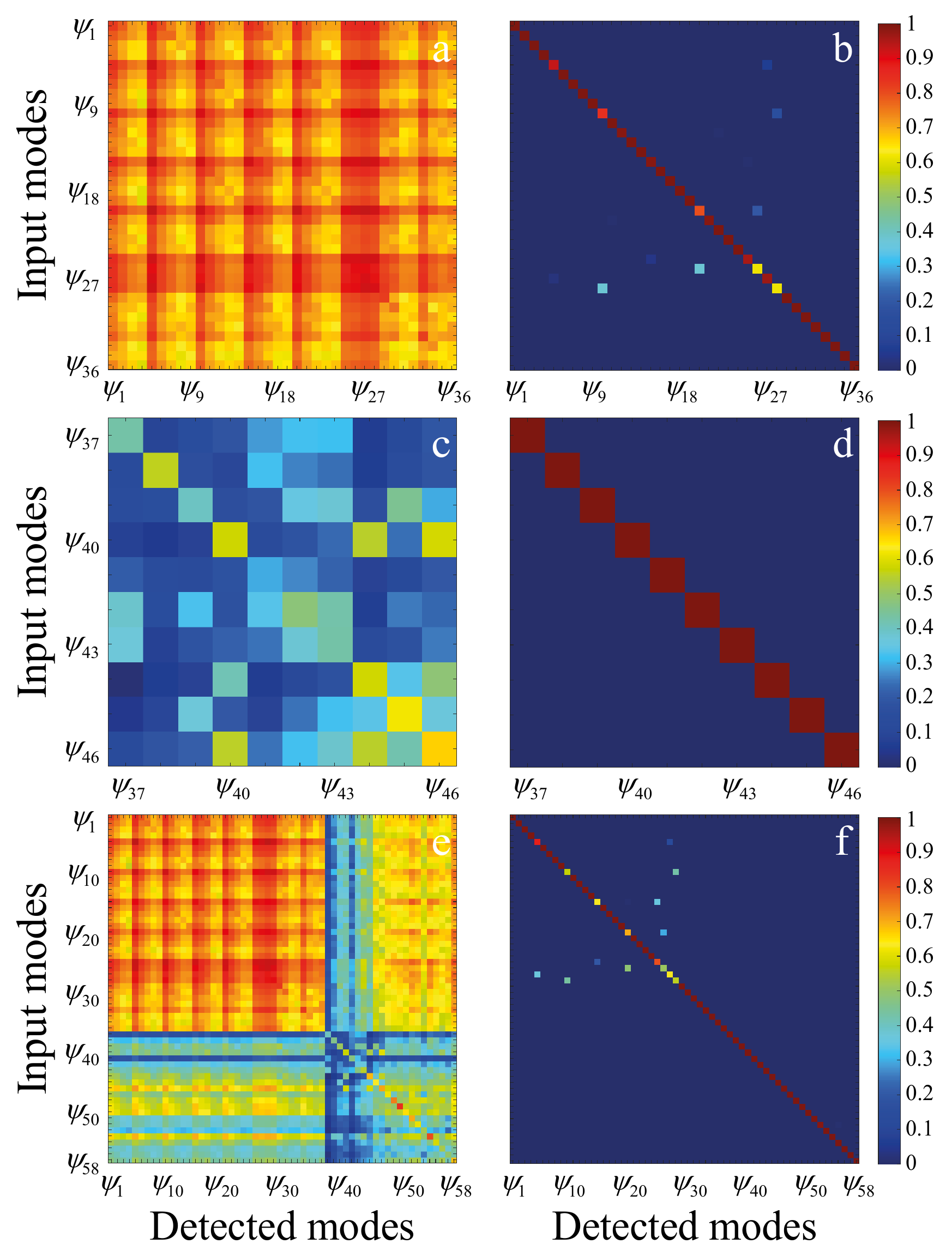}
\caption{\label{fig:Confusion} \textbf{Spatial mode identification: maximum likelihood vs trained artificial neural networks.} \textbf{a,b} show the confusion matrices for individual (single) modes; \textbf{c,d} LG superpositions, and \textbf{e,f} non-orthogonal superpositions. The left-hand column shows the confusion matrix using the maximum likelihood, defined in Eq. (\ref{eq:likelihood}). Note that, in all cases, the identification of spatial modes using maximum likelihood as metric is impossible. The right-hand column displays the confusion matrices for our trained neural networks. The diagonal elements correspond to correctly identified spatial modes, while off-diagonal elements denote misclassifications.}
\end{figure}

\begin{figure}[t!]
\includegraphics[width=1\linewidth]{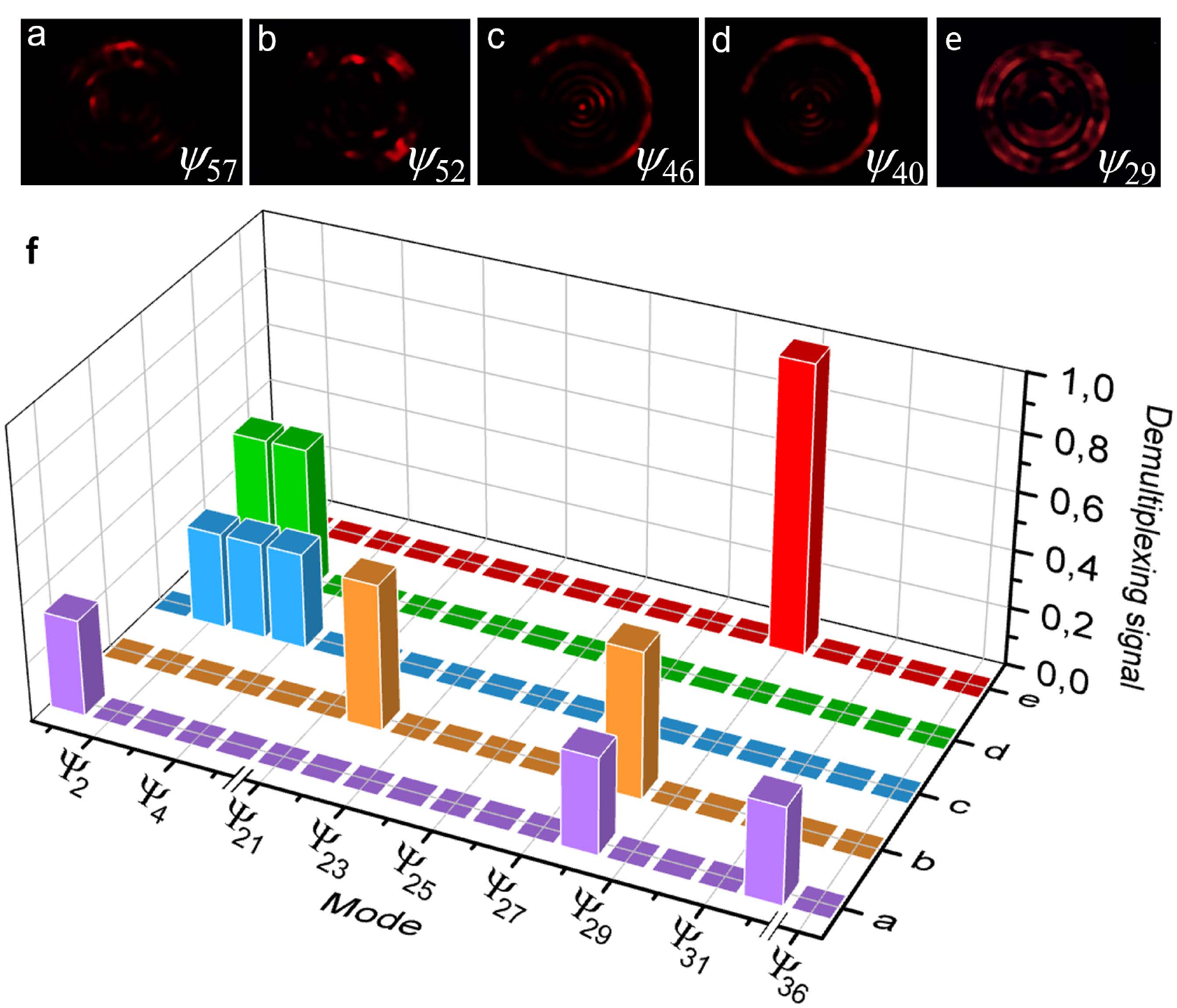}
\caption{\label{fig:Demultiplexing} \textbf{Machine-learning-enabled spatial-mode demultiplexing.} \textbf{a} and \textbf{b} Distorted intensity distributions of non-orthogonal mode superpositions, namely $\psi_{57}=LG_{03}+V_{-2}+B_{0,0.09}$, and $\psi_{52}=LG_{-25}+V_{-2}$. \textbf{c} and \textbf{d} Distorted intensity distributions of superpositions comprising two and three LG modes, $\psi_{46}=LG_{05}+LG_{07}+LG_{09}$ and $\psi_{40}=LG_{05}+LG_{07}$. \textbf{e} Distorted intensity distribution of an individual mode $\psi_{29}=B_{0,0.03}$. \textcolor{black}{\textbf{f} Bar plot shows the OAM spectrum obtained by our smart machine-vision device from each detected intensity distribution. In case \textbf{a}, represented by purple bars, the modes $\psi_{1}=LG_{03}$, $\psi_{28}=V_{-2}$, and $\psi_{32}=B_{0,0.09}$ contribute to one-third of the coherent superposition labeled as $\psi_{57}$. Similarly, in case \textbf{c}, shown by blue bars, the mode $\psi_{46}$ comprises the the fields labeled as $\psi_{2}=LG_{05}$, $\psi_{3}=LG_{07}$, and $\psi_{4}=LG_{09}$. For case \textbf{b}, indicated by orange bars, the modes $\psi_{22}=LG_{-25}$ and $\psi_{28}=V_{-2}$ contribute equally to the detected $\psi_{52}$ mode, each with a weight of one-half. The equivalent situation occurs for the case \textbf{d}, represented by green bars, where the modes $\psi_{2}=LG_{05}$ and $\psi_{3}=LG_{07}$ make up for $\psi_{40}$.  Finally, the contribution of the mode $\psi_{29}=B_{0,0.03}$ is unitary, as represented by the red bar.}}
\end{figure}

We have trained three different artificial neural networks to demonstrate the capabilities of our smart camera for recognizing and reconstructing distorted spatial modes. Our first neural network, comprising ten neurons in the single hidden layer, recognizes individual modes, including LG modes with different azimuthal and radial parameters, Vortex beams, and Bessel modes with different orders and radial wave vectors. Notably, this neural network can identify thirty-six classes, as depicted in the blue outer circle in Fig. \ref{fig:Arreglo}b. Our second neural network, which also includes ten neurons in its hidden layer, is designed to demultiplex ten different LG superpositions, where five output classes correspond to superpositions of two LG modes and five to superpositions of three LG modes. These classes are shown in the middle orange circle of Fig. \ref{fig:Arreglo}b. The demultiplexing of non-orthogonal mode superpositions (i.e. those composed by Bessel, Vortex, and LG modes) is achieved by making use of a third neural network, which uses fifteen neurons in its hidden layer to complete such task. The inner green circle of Fig. \ref{fig:Arreglo}b shows the twelve classes corresponding to the superpositions of two and three non-orthogonal modes. Importantly, this third neural network cannot only demultiplex superpositions of non-orthogonal modes but also LG superpositions and individual modes, thus leading to the identification and demultiplexing of the fifty-eight different classes of spatial modes, see Appendix B for details on the modes, and their corresponding labeling, used below. 

To provide a point of comparison between commonly used identification methods and our smart vision device, we have calculated the cross-talk of maximum likelihood for individual modes, LG superpositions, and non-orthogonal superpositions. The left-hand column of Fig. \ref{fig:Confusion} shows the results obtained in each case. Note that, in the cross-talk matrix, the diagonal elements represent the conditional probabilities among the input (noise-free) and detected (distorted) modes that are correctly matched. In other words, the diagonal elements give the probability of correctly identifying a distorted spatial mode; whereas the off-diagonal elements represent the times the algorithm fails to correctly perform the classification task. We evaluate the maximum likelihood using the expression: 
\begin{equation}\label{eq:likelihood}
\Omega=\frac{\left[\int I_{\text{in}}^{1/2}(\mathbf{x})I_{\text{det}}^{1/2}(\mathbf{x})d\mathbf{x}\right]^2}{\left[\int I_{\text{in}}(\mathbf{x})d\mathbf{x}\right]\left[\int I_{\text{det}}(\mathbf{x})d\mathbf{x}\right]},
\end{equation}
where $\mathbf{x}=(x,y)$ denotes the transverse coordinates in the detection plane. Note that when $\Omega=1$, similarity between the input (noise-free), $I_{\text{in}}(\mathbf{x})$, and the detected (distorted) spatial modes, $I_{\text{det}}(\mathbf{x})$, is perfect. 

In all cases, the identification and reconstruction of spatial modes is impossible using the maximum likelihood. Notably, these results contrast with the confusion matrices of the trained neural network, shown on the right-hand column of Fig. \ref{fig:Confusion}. The neural network designed to reconstruct individual modes achieved an overall accuracy of 95\%. By examining the confusion matrix depicted in Fig. \ref{fig:Confusion}b, the off-diagonal elements make the misclassifications evident. Interestingly, a significant number of these misclassifications are observed between LG beams with a radial mode number $p=0$ and vortex beams. This can be attributed to the high similarity in the intensity distributions of these modes. However, the identification is perfect for our neural network that reconstructs LG superpositions. In this case, the confusion matrix does not have off-diagonal elements, as shown in Fig. \ref{fig:Confusion}d. Figure \ref{fig:Confusion}f shows the complete confusion matrix associated to the neural network trained to identify individual modes, orthogonal superpositions, and non-orthogonal superpositions. After the training stage, an overall identification accuracy of 95\% was achieved. This result indicates the effectiveness of the neural network in accurately classifying and distinguishing between different types of spatial modes. The high accuracy rate reflects the successful learning and generalization capabilities of the trained network. Remarkably, Fig.~\ref{fig:Demultiplexing} shows the possibility of using our smart camera for performing mode demultiplexing of different superposition modes. These results demonstrate the potential of our smart machine-vision device for identifying and reconstructing OAM spectra. We have tested our smart camera for non-orthogonal modes superpositions, as shown in Figs.~\ref{fig:Demultiplexing}a and \ref{fig:Demultiplexing}b.  We have also considered coherent superpositions of two and three LG modes (as depicted in Figs.~\ref{fig:Demultiplexing}c and \ref{fig:Demultiplexing}d), as well as individual modes (see Fig.~\ref{fig:Demultiplexing}e). In all cases, our smart camera can reconstruct the OAM spectrum of the experimentally-detected, distorted intensity distributions. \textcolor{black}{The obtained OAM spectrum for each detected intensity distribution is presented in the bar plot of Fig.~\ref{fig:Demultiplexing}f}. It is worth pointing out that our device capabilities for identifying and demultiplexing superpositions of non-orthogonal modes allows for increasing the bandwidth of communication systems \cite{wan2022divergence}, while maintaining a small divergence of the modes, see details in Appendix B. 

Finally, we remark that one of the significant advantages of our neural network architectures is their simplicity.  This not only reduces the training time but also makes them well-suited for deployment in the Raspberry Pi systems. Importantly, all our networks have no more than fifteen neurons in the hidden layer, resulting in a compact and lightweight structure. Despite its simplicity, our neural networks demonstrate remarkable performance in accurately identifying various OAM-carrying modes.

\section{Conclusions}
\label{sec:conclusion}
We have shown that a compact image sensor can itself behave as an artificial neural network that simultaneously identify, reconstruct and demultiplex complex spatially-structured optical fields. We have demonstrated the capability of the smart device in the reconstruction of distorted OAM-carrying modes, as well as their orthogonal, and non-orthogonal superpositions with an accuracy of 95$\%$. Our device includes a model of machine learning that comprises an easy-to-implement two-layer artificial neural network that makes it suitable for deploying in compact electronic devices. Our proposed smart machine-vision device offers a promising avenue towards the development of low-power-consumption smart, optical communication systems.

\section*{Acknowledgments}
\noindent M.A.Q.-J. thankfully acknowledges financial support by CONAHCyT under the Project CF-2023-I-1496 and by DGAPA-UNAM under the Project UNAM-PAPIIT TA101023. J.D.H.-M. thankfully acknowledges financial support by CONAHCyT. S.-H.D. acknowledges the partial support of project 20230316-SIP-IPN. C.Y. and O.S.M.-L. thank the U.S. Department of Energy (DOE), Office of Science (SC), Program: Nuclear Physics for support under the NP-QIS grant: DE-SC0023694. R.J.L.-M. thankfully acknowledges financial support by DGAPA-UNAM under the project UNAM-PAPIIT IN101623.

\section*{Competing interests}
\noindent The authors declare no competing interests.

\section*{Data Availability}
\noindent The data that support the findings of this study are available from the corresponding authors upon reasonable request.


\appendix


\section*{Appendix A. Programming of the smart camera}

We provide a functional description of the codes embedded in the Raspberry Pi for controlling the camera and executing the trained neural network.  Raspberry is a micro-computer that uses an optimized UNIX/Debian-based Raspian operating system that runs on a Broadcom BCM2835 SoC processor. To enable programming on the Raspberry Pi, a USB keyboard and mouse can be connected to the circuit board. These peripherals provide a convenient input method for interacting with the micro-computer and allow for generating scripts using different programming languages. Moreover, the circuit board implements a wireless module for Internet connection. As mentioned in the main text, we establish a connection between the Raspberry Pi camera module and the Raspberry Pi using the serial port. In order to configure this module and program the neural network, we employ the Python language within the open-source Spyder IDE environment. To collect images for both the training and testing phases, we develop a script that utilizes the \textit{Picamera} package \cite{scikit-learn}. This script allows us to configure the camera, set up the image acquisition process, and specify the storage path for the captured images. We set the resolution of the camera to 875 x 656 pixels and the shutter speed to 6000. To reduce the volume of data stored on the internal memory of the Raspberry Pi, the acquired images by the camera are saved in a portable USB memory.

\begin{table*}[t!]
	\centering
	\caption{\bf Spatial mode catalogue.}
	\begin{tabular}{lll}
		\hline \hline 
	\textbf{Modes}	&  &  \textbf{Single modes} \\
		\hline  &  & \vspace{-0.6em}  \\
		\textbf{Bessel beams} &  & $B_{0,0.03};B_{0,0.05};B_{0,0.07};B_{0,0.09};B_{1,0.03};B_{1,0.05};B_{1,0.07};B_{1,0.09}$ \vspace{0.4em}  \\
		\textbf{Vortex beams} &  & $V_{-2};V_{-1};V_{1};V_{2}$ \vspace{0.4em}             \\
		 &  & $LG_{-20};LG_{-23};LG_{-25};LG_{-27};LG_{-29};LG_{20};LG_{23};LG_{25};LG_{27};LG_{29};LG_{03};LG_{05}$ \vspace{-0.5em}  \\ \textbf{LG beams} & & \vspace{-0.5em} \\&  & $LG_{-10};LG_{-13};LG_{-15};LG_{-17};LG_{-19};LG_{10};LG_{13};LG_{15};LG_{17};LG_{19};LG_{07};LG_{09}$  \vspace{0.4em} \\  
		\hline  \hline
       \textbf{No. of modes} &  & \textbf{LG superpositions}  \\
		\hline  & & \vspace{-0.6em} \\ \textbf{Two}  &  & $LG_{00}+LG_{03};LG_{00}+LG_{05};LG_{03}+LG_{09};LG_{05}+LG_{07};LG_{07}+LG_{09}$ \vspace{0.4em} \\   &   & $LG_{00}+LG_{03}+LG_{07};LG_{00}+LG_{03}+LG_{09};LG_{03}+LG_{05}+LG_{07}$ \vspace{-0.5em}   \\ \textbf{Three}  & & \vspace{-0.5em}   \\ &  & $LG_{03}+LG_{05}+LG_{09};LG_{05}+LG_{07}+LG_{09}$  \vspace{0.4em} \\
        \hline  \hline 
       \textbf{No. of modes} &  & \textbf{Non-orthogonal superpositions }  \\  \hline  & & \vspace{-0.6em} \\ \textbf{Two}  &  & $V_{-2}+B_{0,0.03};V_{1}+B_{0,0.05};V_{-2}+LG_{-25};V_{1}+LG_{29};B_{0,0.07}+LG_{-23};B_{0,0.09}+LG_{29}$ \vspace{0.4em} \\    &   & $V_{-2}+B_{0,0.09}+LG_{03};V_{1}+B_{0,0.09}+LG_{03};V_{1}+B_{0,0.03}+LG_{-15}$ \vspace{-0.5em}   \\ \textbf{Three} & & \vspace{-0.5em}   \\ &  & $V_{1}+B_{0,0.09}+LG_{10};V_{-2}+B_{0,0.05}+LG_{29};V_{-2}+B_{0,0.09}+LG_{-23}$   \vspace{0.4em} \\
        \hline  \hline
	\end{tabular}
	\label{tab:Tab1}
\end{table*}

We developed a second script that comprises the image processing, i.e., the conversion from RGB to grayscale, the cropping of the image, the reduction of the image size through a down-sampling process, and the rearrangement of the matrix into a column vector. This script also facilitates the generation of an input data matrix, which comprises all the observations, and a target matrix that specifies the class to which each observation belongs. By executing this script, the necessary pre-processing steps are performed, resulting in organized and formatted data for the training and testing phases of the neural network. 

The third script comprises the core smart camera algorithm, incorporating the acquisition of spatial mode intensity distribution, pre-processing of the captured image, and the implementation of the trained neural network for classification purposes. It is important to highlight that the codes developed for our smart camera, particularly for pre-processing and the neural network, are entirely custom-made and do not rely on pre-existing package modules. This highlights the originality and uniqueness of the implemented codebase. The codes embedded in the Raspberry Pi for controlling the camera and executing the trained neural network are available from GitHub repository at \textcolor{blue}{https://github.com/mquirozj/Smart-camera-for-spatial-mode-reconstruction}.


\section*{Appendix B. Spatial mode catalogue and divergence analysis} 

\subsection{Spatial mode catalogue}
\begin{figure*}
\includegraphics[width=1\linewidth]{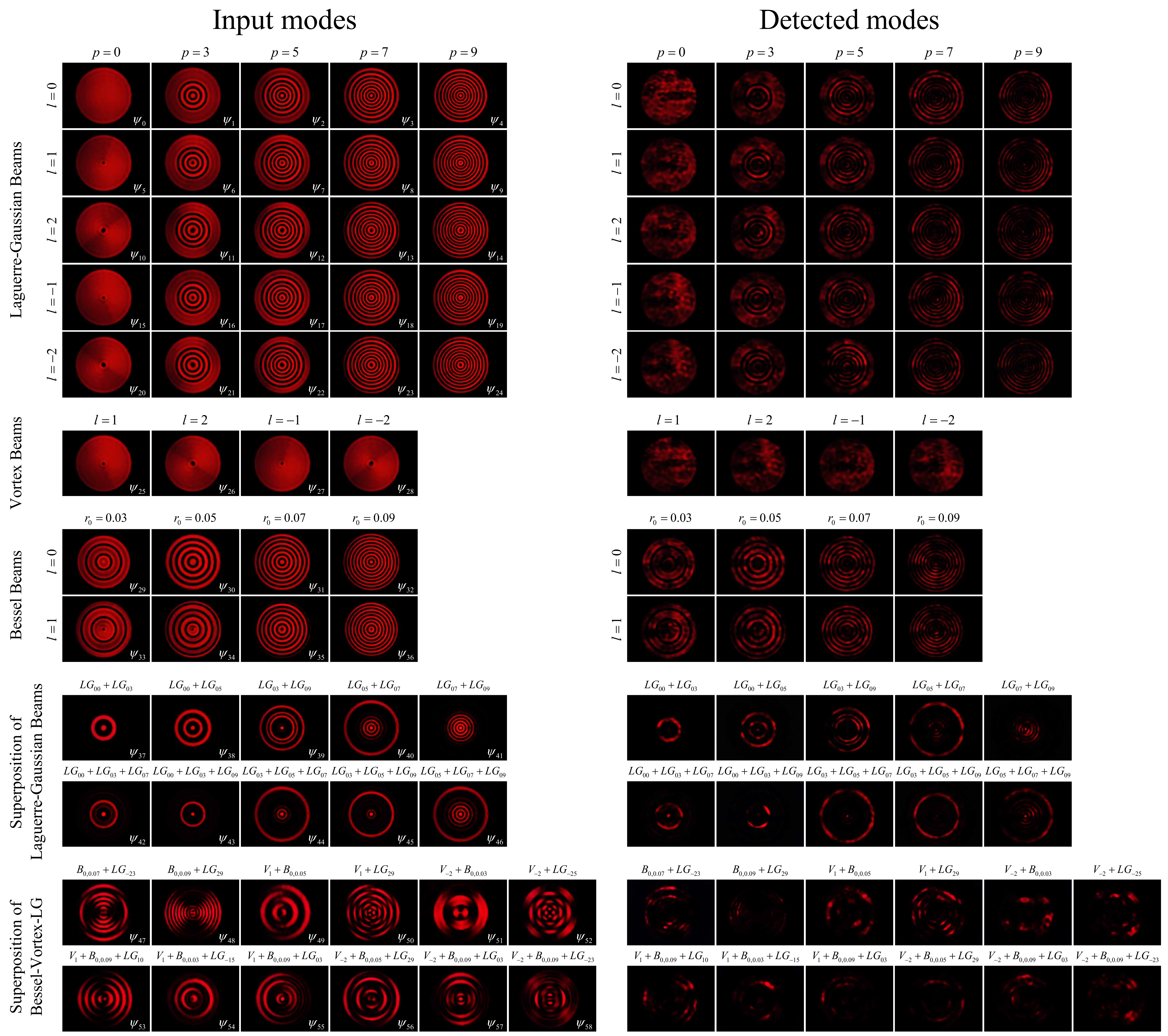}
\caption{\textbf{Intensity distribution of Laguerre-Gaussian, Vortex, and Bessel beams.} The spatial modes on the left-hand side (Input modes) correspond to the dispersion-free signal captured by the smart camera, while the modes on the right-hand side (Detected modes) are affected by dispersion.}
\label{fig:One_Mode}
\end{figure*}

Figure~\ref{fig:One_Mode} shows the catalogue of spatial modes used to train and test our neural networks. The catalogue has been prepared on a basis of 36 OAM-carrying individual modes, 10 LG superposition modes and 12 non-orthogonal superposition modes. Withing the 36 single modes, we generate 8 Bessel beams, 4 Vortex beams and 24 LG beams. Moreover, for the LG and non-orthogonal superpositions we randomly select from the 36 single modes to generate superpositions of two and three fields, see Table~\ref{tab:Tab1} for details. In the images shown on the left-hand side of Figure~\ref{fig:One_Mode} we see the input modes captured by the smart camera without noise in the signal. The intensity profiles on the right-hand side correspond to the detected modes by our smart device in the presence of noise, introduced by the ground glass. It should be noted that we have arbitrarily chosen only one detected mode with noise to report in this figure; however, the training of the neural networks is performed using 100 data images, captured for each of the classes.

\subsection{Divergence of spatial modes}
One of the main challenges in communication systems and information coding by means of spatial-mode demultiplexing is the significant degeneracy of the beam due to power loss and high divergence of the propagating fields. In this sense, our non-orthogonal mode superpositions proposal represents an improvement over the challenge of divergence. This type of mode combinations shows a considerably low variation of the beam spot diameter. More specifically, in the following, we show that the divergence angle of the non-orthogonal superposition modes has a lower growth rate compared to the divergence of Vortex and LG beam modes.

\begin{figure}[t!]
\includegraphics[width=1\linewidth]{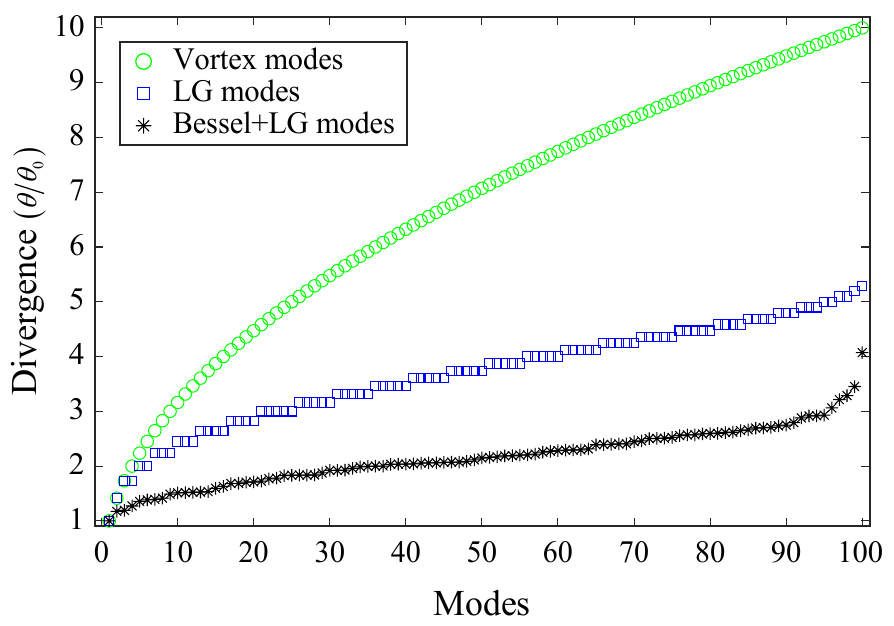}
\caption{\textbf{Divergence of OAM-carrying spatial modes}. Normalized divergence for Vortex modes in green circles, LG modes in blue squares, and non-orthogonal mode superpositions of Bessel+LG beams in black asterisks. We compare the divergence of 100 modes (see main text for details) in each case.}
\label{fig:divergence} 
\end{figure}

Figure~\ref{fig:divergence} shows the curves for the normalized divergence of Vortex, LG, and Bessel+LG superposition modes. We consider 100 modes in each case, where we normalize to the divergence angle of a fundamental Gaussian beam $\theta_{0}$. For this, we use the second moment of intensity of the beam to calculate the size of the spot diameter as \cite{Phillips:83,Willner_2016,wan2022divergence},
\begin{eqnarray}
    D_z=2\sqrt{ \frac{2 \int_{0}^{2\pi} \int_{0}^{\infty}{r^2 I_{z}(r,\phi) rdrd\phi }}{\int_{0}^{2\pi}\int_{0}^{\infty}{I_{z}(r,\phi) rdrd\phi }} },
\end{eqnarray}
where $I_{z}(r,\phi)$ denotes the intensity profile of the beam at a propagation distance $z$. We obtain the value of the divergence angle by calculating the beam diameter at $z=0$, which we identify as the initial diameter $D_i$, and the final diameter $D_f$ corresponding to a specific propagation distance $z=z_{f}$. Following previous authors \cite{wan2022divergence}, we propagate the modes twice their Rayleigh distance $\left( z_f= 2z_R \right)$. By using the above mentioned parameters, we can then find the beam divergence as
\begin{eqnarray}
    \theta=2 \arctan \left( {\frac{D_f-D_i}{4z_R}} \right).
\end{eqnarray}

\noindent It is worth highlighting that the comparison of beam divergence in Fig.~\ref{fig:divergence} corresponds to modes generated under the following configurations:
for Vortex modes we have $l=0,...,99$, for LG modes we have considered $p=0,...,9$ and $l=0,...,9$; finally for non-orthogonal superpositions we have chosen 100 combinations Bessel+LG generated in the basis $\left\{ B_{0,0.03},B_{0,0.05},B_{0,0.07},B_{0,0.09} \right\}$ and all possible combinations for $LG_{lp}$ taking $l=0,1,2,3,4$ and $p=0,1,2,3,4$. Note that the angle of divergence for the mixtures of Bessel+LG modes, data in black asterisks in Fig.~\ref{fig:divergence}, has a moderate growth compared to the large values in green circles and blue squares corresponding to Vortex and LG modes, respectively. In other words, the modes of the non-orthogonal superpositions exhibit less divergence than the other OAM-carrying individual modes. Moreover, their divergence values maintain a fairly constant behavior. This is clearly observed, for instance, between modes 30 and 60, where the normalized divergence stays around 2. These results show that non-orthogonal mode superpositions and smart machine-vision devices are a promising route toward transmission, reconstruction, and demultiplexing of optical information systems.

\bibliography{biblio}

\end{document}